\documentclass[aps,pra,showpacs,
twocolumn,
superscriptaddress
]{revtex4-1}

\usepackage[T1]{fontenc}
\usepackage{lmodern}
\usepackage[pdftex]{graphicx}
\usepackage{amsmath}

\usepackage{xcolor}
\usepackage[%
  colorlinks=true,
  urlcolor=blue,
  linkcolor=blue,
  citecolor=blue
]{hyperref}

\usepackage{amssymb,bm}
\usepackage{braket}

\begin{document}
\title{Magneto-optical trapping of optically pumped metastable europium%
}
\author{Ryotaro~Inoue}
\affiliation{Department of Physics, Graduate School of Science, Tokyo Institute of Technology, 2-12-1 O-okayama, Meguro-ku, Tokyo 152-8550 Japan}
\author{Yuki~Miyazawa}
\affiliation{Department of Physics, Graduate School of Science, Tokyo Institute of Technology, 2-12-1 O-okayama, Meguro-ku, Tokyo 152-8550 Japan}
\author{Mikio~Kozuma}
\affiliation{Department of Physics, Graduate School of Science, Tokyo Institute of Technology, 2-12-1 O-okayama, Meguro-ku, Tokyo 152-8550 Japan}

\begin{abstract}

We demonstrate laser cooling and magneto-optical trapping of europium.
The atoms are optically pumped to a metastable state and then loaded from an atomic-beam source via conventional Zeeman slowing and magneto-optical trapping techniques using a $J=13/2\leftrightarrow J=15/2$ quasi-cyclic transition.  
The trapped populations contained up to $1\times 10^7$ atoms, and a two-body loss rate is estimated as $1\times10^{-10}\,\mathrm{cm^3/s}$ from the non-exponential loss of atoms at high densities.
We also observed leakage out of the quasi-cyclic transition to the two metastable states with $J=9/2$ and $11/2$, which is adequate to pump the laser-cooled atoms back to the $J=7/2$ ground state.
 
\end{abstract}
\maketitle

Magneto-optical traps (MOTs) for atoms have become standard tools in atomic physics, enabling the investigations of fundamental and applied physics in diverse scientific topics such as degenerate quantum gases \cite{Bloch2008}, quantum information processing \cite{Lukin2003,Saffman2010}, and optical frequency standards \cite{Derevianko2011}.
Laser cooling with optical transitions in a nearly closed subset of atomic states is required to operate a MOT.
Except for ytterbium, the lanthanides possess complex energy structures owing to the presence of open f-shells. Forming a MOT with an open f-shell-lanthanide was first demonstrated in 2006 using erbium \cite{McClelland2006}. Demonstrations of MOTs with dysprosium \cite{Youn2010}, thulium \cite{Sukachev2010}, and holmium \cite{Miao2014} followed a few years later. Exploring the lanthanides is motivated by several topics in current research areas.
Examples include dipolar phenomena relying on magnetic dipole-dipole interactions \cite{Lahaye2008,Ferrier-Barbut2016,Chomaz2016a,Kadau2016} that capitalize on the large magnetic dipole moments of the lanthanides as well as quantum information processing using collective encoding of multi qubit quantum registers \cite{Saffman2016} that depend on the large spin degrees of freedom of these elements.

Herein, we report the formation of a MOT using europium (Eu), a lanthanide possessing two stable bosonic isotopes:  ${}^{151}\text{Eu}$ (comprising $48\,\mathrm{\%}$ of the natural abundance of this element) and ${}^{153}\text{Eu}$ ($52\,\mathrm{\%}$).
Both isotopes have nuclear spins $I=5/2$ and large dipole moments $\mu=7\,\text{Bohr magnetons}\,(7\mu_B)$.
The Eu atom has a spherically symmetric ${}^{8}\mathrm{S}_{7/2}$ electronic ground state in contrast to the other dipolar lanthanides. The large electronic orbital angular momentum of the ground state provides strong anisotropy, which leads to a rich behavior of the interactions between the atoms \cite{Maier2015,Nathaniel2016}. The highly symmetric electronic ground state of Eu reduces the density of magnetic-field-induced Feshbach resonances \cite{Suleimanov2010}.
In addition to the dimer $\mathrm{Eu}_2$ \cite{Alexei2009}, Eu forms heteronuclear molecules with alkaline or alkaline-earth atoms, which possess both large electric and large magnetic dipole moments. These molecules are promising candidate systems for quantum simulations of many-body physics \cite{carr2009,Tomza2014}.

The hyperfine structure of Eu enables us to control the scattering length using RF fields \cite{Papoular2010,Hanna2010} without applying static magnetic fields.
In addition, the hyperfine state with the largest magnetic-dipole moment in the electronic ground state corresponds to the lowest-energy state in the hyperfine manifold.
These features are of interest for exploring ground-state phases in dipolar spinor Bose gases \cite{Yi2006,Kawaguchi2006,Takahashi2007,Pasquiou2011}.
Manipulation of contact interactions at low magnetic fields ($\lesssim 1\,\mathrm{mG}$ \cite{Pasquiou2011}), such that Zeeman effects do not obscure the dipolar effects, becomes crucial for experimental investigations of the true quantum ground state of the system.

\begin{figure*}[btp]
\includegraphics{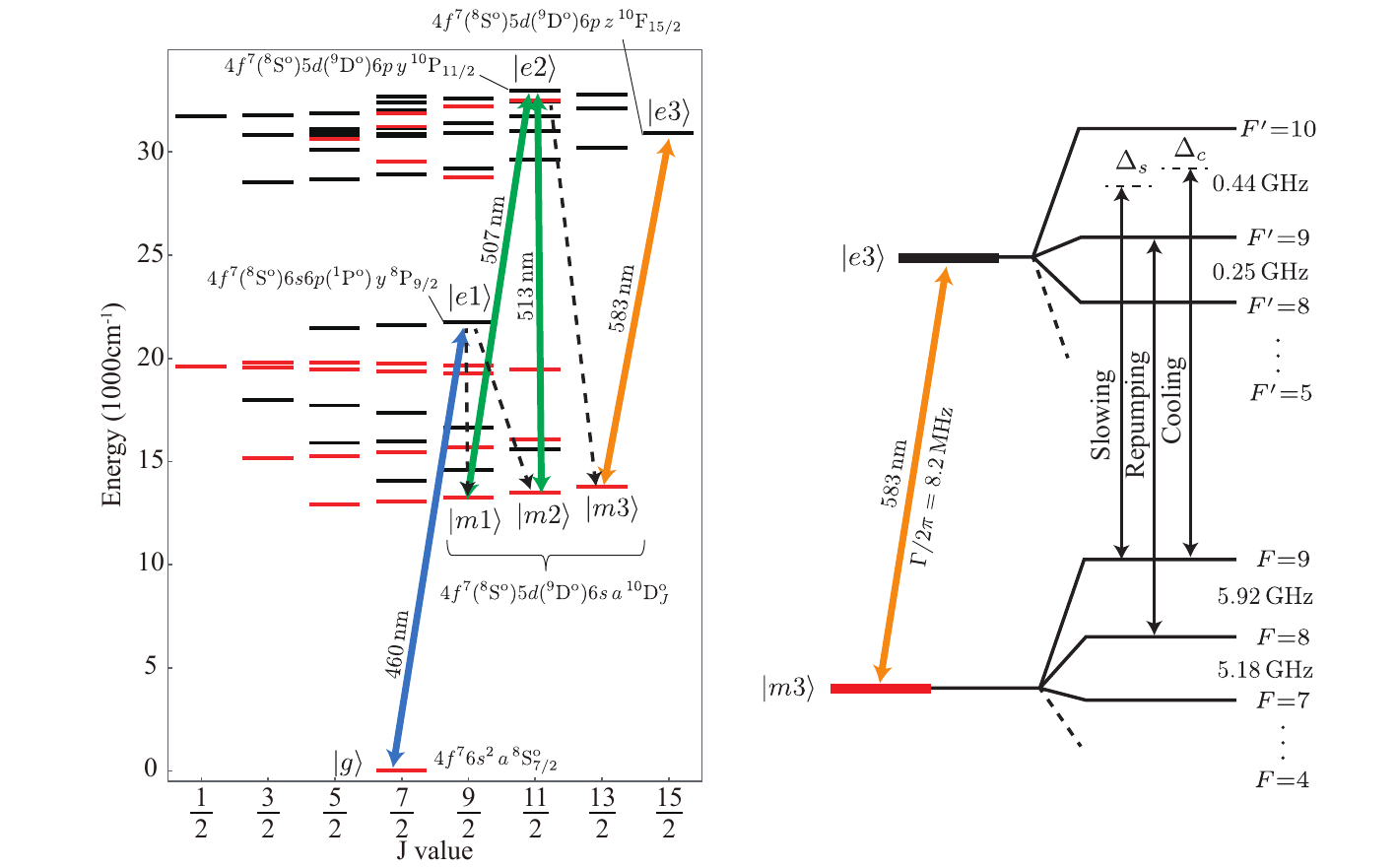}
\caption{(Color online) The energy levels of Eu \cite{Martin1978,Hartog2002} showing the pumping and laser-cooling transitions.
We have plotted the levels up to the energy of $32948.41\,\mathrm{cm^{-1}}$, which corresponds to the uppermost level $\ket{e2}$ in our experiment, while some levels are omitted because they have not been fully interpreted \cite{Martin1978}.
Odd-parity states are indicated by gray lines (red online). 
Bold arrows indicate laser-driven transitions, and dashed arrows show spontaneous decay channels relevant to our optical pumping procedure. Although the excited states $\ket{e1}$, $\ket{e2}$, and $\ket{e3}$ have additional decay channels, here we omit the decays to the other states for clarity. The level diagram on the right shows the hyperfine level structures of the $583$-nm transition, which were calculated from the A and B coefficients for ${}^{151}\mathrm{Eu}$ \cite{Brand1981}. The transition $F=9\leftrightarrow F'=10$ is used for the laser-cooling and the Zeeman slowing, while the transition $F=8\leftrightarrow F'=9$ is used for the repumping.
}
\label{fig:energylevels}
\end{figure*}

An energy level diagram of neutral $\mathrm{Eu}$ \cite{Martin1978,Hartog2002} is shown in Fig.~\ref{fig:energylevels}. The ground state $\ket{g}$ has angular momentum $J=7/2$ with odd parity. The transition to the state, labeled by $\ket{e1}$, is the only dipole-allowed transition at wavelengths longer than $400\,\mathrm{nm}$ (energies $\le 25000\,\mathrm{cm^{-1}}$) to an even-parity state with $J'=9/2$. This is a strong $J\leftrightarrow J+1$ transition: the transition wavelength is $460\,\mathrm{nm}$ and exhibits a broad linewidth of $27\,\mathrm{MHz}$. However, eleven low-lying odd-parity states exist for $\ket{e1}$.  The lower limit to the probability of leakage of the $\ket{g}\leftrightarrow\ket{e1}$ system was spectroscopically determined to be $1.05(2)\times10^{-3}$ \cite{Miyazawa2017}. This suggests a pessimistic outlook for laser-cooling using this transition; we cannot expect to laser-cool the atoms without using multiple repumping lasers operating at different wavelengths to plug the leaks. A simple alternative is to use the cyclic transition $\ket{m3}\leftrightarrow\ket{e3}$ (with a transition wavelength of $583\,\mathrm{nm}$ and a natural linewidth of $\Gamma/2\pi=8.2\,\mathrm{MHz}$). For laser cooling using this transition, the atoms are first pumped from the ground state $\ket{g}$ to the intermediate states $\ket{m1}$ and $\ket{m2}$ by driving the $\ket{g}\leftrightarrow\ket{e1}$ transition with a pumping beam at $460$-nm wavelength. Adding two additional pumping beams at the wavelengths $507$ and $513\,\mathrm{nm}$ enables us to pump the atoms to $\ket{m3}$, as schematized in Fig.~\ref{fig:energylevels}.

Our Eu atomic beam was produced using an effusive oven operating at $870\,\mathrm{K}$. The atoms emitted from the oven are first pumped to $\ket{m3}$ and then decelerated in a $210$-mm-long Zeeman slower. Our Zeeman slower operates with light that is red-detuned by $\Delta_s/ 2\pi =240\,\mathrm{MHz}$ from the unshifted resonance of the cyclic transition $F=9\leftrightarrow F'=10$. This system allows the deceleration of ${}^{151}\mathrm{Eu}$ atoms in the metastable state $\ket{m_3,\ F=9}$ that have velocities lower than $150\,\mathrm{m/s}$, which corresponds to the lower $7.5\,\mathrm{\%}$ of the velocity distribution. The three-color pumping light beams (at the wavelengths of $460$, $507$, and $513\,\mathrm{nm}$) copropagate along the slowing light beam. The optically pumped and decelerated atoms are fed into the chamber used for the MOT, which has a pressure of $\sim 1\times 10^{-8}\,\mathrm{Pa}$ with the atomic beam running. Our MOT is formed in a quadrupole magnetic field provided by anti-Helmholtz coils and three pairs of counter-propagating, circularly polarized cooling-light beams. The cooling light is also red-detuned by $\Delta_c$ from the $F=9\leftrightarrow F'=10$ cyclic transition. On-resonant repumping light tuned to $F=8 \leftrightarrow F'=9$ overlaps with all the cooling beams. The beams have Gaussian waists ($1/e^2$ intensity radii) of $11\,\mathrm{mm}$ and are truncated by a circular aperture of $21.6$-mm-diameter.

\begin{figure}[tbp]
\includegraphics{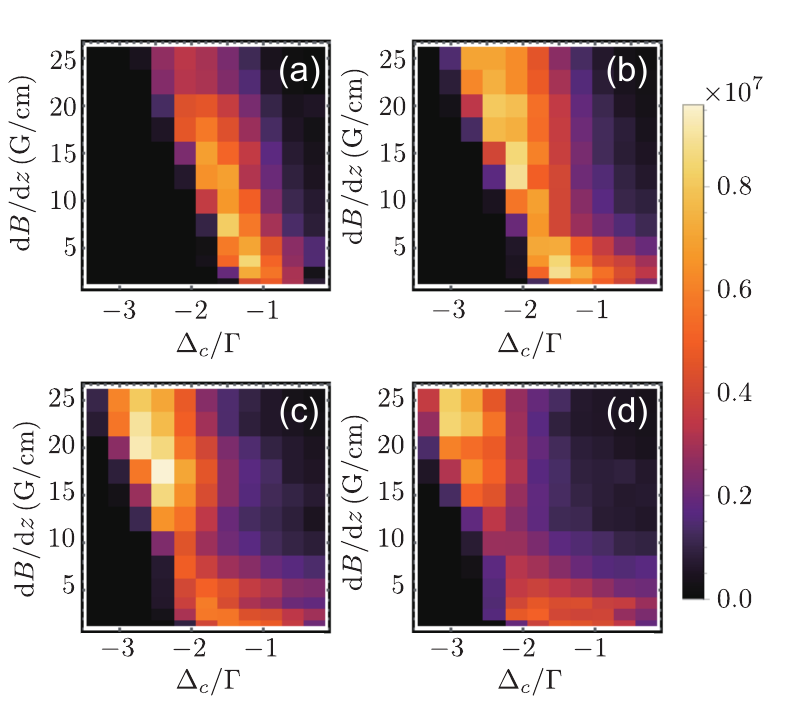}
\caption{(Color online) The number of trapped atoms as a function of the magnetic-field gradient $\mathrm{d}B/\mathrm{d}z$ and the detuning of the cooling light ($\Delta_c/\Gamma$).
Panels (a)-(d) correspond to the total cooling-light powers $12\,\mathrm{mW}$(a), $24\,\mathrm{mW}$(b), $36\,\mathrm{mW}$(c), and $48\,\mathrm{mW}$(d).}
\label{fig:numofatoms}
\end{figure}

The total number of trapped atoms are determined using an absorption imaging technique with a circularly polarized probe pulse tuned to the $F=9\leftrightarrow F'=10$ transition. After atom loading, we turned off all the light beams and the quadrupole magnetic field and applied a weak magnetic field of $\sim1\,\mathrm{G}$ along a given direction. Irradiation with a two-color, $\sigma^+$-polarized light pulse tuned to $F=9 \leftrightarrow F'=9$ and $F=8\leftrightarrow F'=9$ enables us to pump the atoms into the dark state $\ket{F=9,M_F=9}$, where $M_F$ is a magnetic quantum number. By rotating the magnetic-field either to be parallel or antiparallel to the propagation direction of the probe pulse, we confirm that the degree of spin polarization is sufficient to determine the number of atoms assuming that the atomic ensemble is fully polarized. In addition, we find that the spin polarization of the atoms varies in a complicated manner with the parameters used for the MOT. Consequently, the effective saturation intensity for the multilevel system $F=9\leftrightarrow F'=10$ is not constant, which makes the quantitative determination of excitation fraction difficult.  
Figure~\ref{fig:numofatoms} summarizes the number of trapped atoms as a function of the total power $P$ of the six cooling beams, the detuning $\Delta_c/\Gamma$ of the cooling light, and the magnetic-field gradient $\mathrm{d}B/\mathrm{d}z$ along the symmetry axis $z$. With the optimized parameters $\Delta_c/\Gamma = -2.4$, $P=36\,\mathrm{mW}$, and $\mathrm{d}B/\mathrm{d}z=17.5\,\mathrm{G/cm}$, we successfully loaded a MOT with $1\times10^7$ atoms. 
The power of $36\,\mathrm{mW}$ corresponds to a total intensity $I=1.3I_s$, where $I_s\simeq 2.7\times 5.4\,\mathrm{mW/cm^2}$ is the conventional saturation intensity; the factor of $2.7$ accounts for averaging over equally populated Zeeman sublevels and for all kinds of polarization of the light field in the MOT.
Simplified Doppler-cooling theory \cite{Lett1989} predicts the MOT temperature $T_D$ for arbitrary $I$ and $\Delta_c$ as:
\begin{align}\label{eq:dopplertemperature}
T_D=\frac{\hbar\Gamma}{4k_B}\frac{1+I/I_s+(2\Delta_c/\Gamma)^2}{2|\Delta_c|/\Gamma}.
\end{align}
This temperature has a minimum value (Doppler-cooling limit, $210\,\mathrm{\mu K}$ for the transition at $583\,\mathrm{nm}$) in the limit of $I/I_0\ll 1$ with $2\Delta_c/\Gamma=-1$.
In temperature determination using ballistic expansion, we observed a dual-component gas comprising a cold core surrounded by a hot shell, which resembles the structures reported in a MOT for dysprosium \cite{Youn2010,Lu2010}. For the optimal parameters mentioned above, the temperatures of the core and the shell are estimated to be $230(21)\,\mathrm{\mu K}$ and $1.1(1)\,\mathrm{mK}$, respectively, while the Doppler-cooling temperature $T_D$ is calculated as $\sim 500\,\mathrm{\mu K}$ for this parameter.
Much lower temperatures have been obtained with smaller magnetic-field gradients.
For example, we observed a core temperature of $41(4)\,\mathrm{\mu K}$ and a shell temperature of $410(80)\,\mathrm{\mu K}$ with $\mathrm{d}B/\mathrm{d}z=5\,\mathrm{G/cm}$ and $\Delta_c/\Gamma = -1.1$.
The number of atoms in the core and the shell are $\sim 50\,\%$ of the total population; however, note that this percentage may contain significant error because herein the atom numbers are determined without using the spin polarizing procedure. 

\begin{figure}[tbp]
\includegraphics{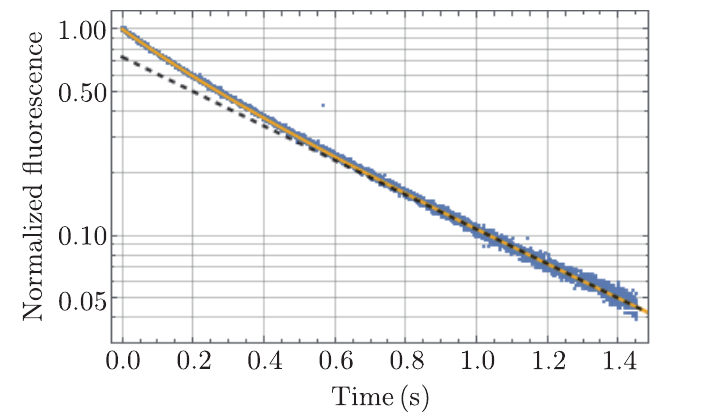}
\caption{(Color online) The observed MOT fluorescence-decay curve for the optimal trapping parameters. The data were obtained by averaging five MOT loading/decay cycles. The solid curve is a fit to the non-exponential decay model given by Eq.~\ref{eq:decay}. For comparison, the dashed line shows a simple exponential fit to the data after $0.75\,\mathrm{s}$.}
\label{fig:decaycurve}
\end{figure}

We also measured the lifetime of atoms in the MOT by interrupting the atom loading to the MOT.
For example, Fig.~\ref{fig:decaycurve} shows the decay curve observed with the optimal MOT parameters, while the pumping laser beam at the wavelength of $460\,\mathrm{nm}$ was turned off to prevent the production of the metastable atoms.
The phenomenological equation for the number $N(t)$ of trapped atoms is as follows: 
\begin{align}\label{eq:rate}
\frac{\mathrm{d}N}{\mathrm{d}t}=-\gamma N-\beta\int\!n^2(\bm{r},t)\mathrm{d}^3r.
\end{align}
Assuming that the density distribution $n(\bm{r},t)$ can be expressed as a Gaussian with a root-mean-square (RMS) size $\sigma_{x,y,z}$ along the three axes, we can rewrite the last term in Eq.~\ref{eq:rate} as $-\beta N^2/V_\text{eff}$, where $V_\text{eff}$ is a volume given by $V_\text{eff}=(2\sqrt{\pi})^{3}\sigma_x\sigma_y\sigma_z$. Herein, we chose the axial direction of the MOT coil as $z$ and the other two orthogonal radial directions as $x$ and $y$. We determined the RMS sizes via absorption imaging with the probe pulse propagating along the $x$ direction, just after releasing the cloud by turning off the light fields for the MOT.
Using the absorption imaging dataset obtained with various MOT-holding durations ($<0.75\,\mathrm{s}$), we found the dimensions to be $\sigma_y=324(8)\,\mathrm{\mu m}$ and $\sigma_z=178(6)\,\mathrm{\mu m}$; the numbers in parentheses here denote the respective standard deviations. We confirmed that these sizes are approximately independent of the MOT-holding durations or the number of atoms remaining in the trap.
Note that, for the following determination of $\beta$, we assume $\sigma_x=\sigma_y$.
The decay of the number of trapped atoms $N$ is obtained by integrating Eq.~\ref{eq:rate}. This gives
\begin{align}\label{eq:decay}
\frac{N}{N_0}=\frac{1}{(1+\gamma_2/\gamma)e^{\gamma t}-\gamma_2/\gamma},
\end{align}
where $N_0$ is the initial number of atoms and $\gamma_2$ is defined as $\beta N_0/V_\text{eff} $. We extracted the coefficients from fitting Eq.~\ref{eq:decay} to the experimental data; this yields $\gamma=1.9\,\mathrm{s^{-1}}$ and $\gamma_2/\gamma=0.55$, which corresponds to $\beta\simeq1\times10^{-10}\,\mathrm{cm^3/s}$. 
This result provides an upper limit to the two-body collision rate constant.
Although the magnitude of $\beta$ obtained for the metastable $\text{Eu}^*$ atoms in the MOT is comparable to that for metastable noble gases in traps without near-resonant light fields \cite{vassen2012}, a penning collision ($\text{Eu}^* + \text{Eu}^* \rightarrow \text{Eu} +\text{Eu}^+ + e^-$) might not occur even for a collision between Eu atoms in a lower state ($\ket{m3}$, $13778.68\,\mathrm{cm^{-1}}$) and an excited state ($\ket{e3}$, $30923.71\,\mathrm{cm^{-1}}$), because the ground-state energy of ionized $\text{Eu}^+$ (${}^9\text{S}_4$, $45734.74\,\mathrm{cm^{-1}}$) is large compared with the energy of the metastable states.
Comparable values of the two-body rate constant can also be found in a system of magnetically trapped $L\not=0$ atoms such as erbium ($3.0\times10^{-10}\,\mathrm{cm^3/s}$ \cite{Connolly2010}) or thulium ($1.1\times10^{-10}\,\mathrm{cm^3/s}$ \cite{Connolly2010}) owing to the long-range electrostatic quadrupole-quadrupole interaction, in contrast to the $L=0$ ground state of europium ($2.5(1.5)\times10^{-13}\,\mathrm{cm^3/s}$; see reference \cite{Kim1997}).
Further investigations will focus on collisional processes among Eu atoms at rest in a magnetic or an optical trap. 

\begin{figure}[btp]
\includegraphics{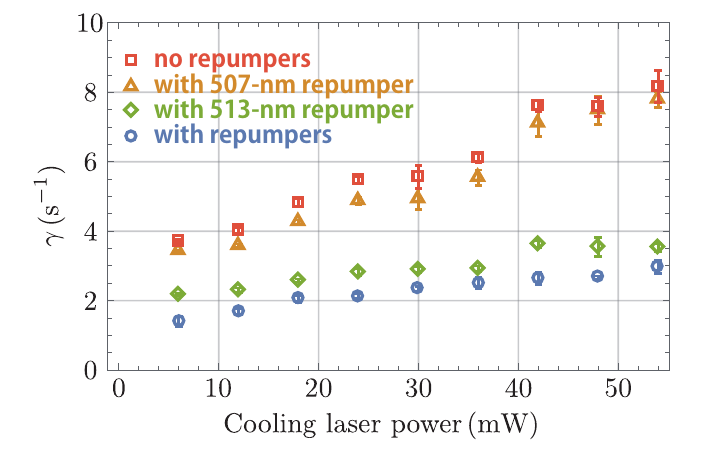}
\caption{(Color online) Dependence of the one-body loss rate $\gamma$ on the repumping lasers or repumpers (at the wavelengths of $507$ and $513\,\mathrm{nm}$) and the total power of the cooling light. The magnetic-field gradient and the detuning of the cooling light were $10\,\mathrm{G/cm}$ and $-0.5\Gamma$, respectively.}
\label{fig:1-body}
\end{figure}

To determine the one-body loss rate $\gamma$, we measured the fluorescence decays by changing the power of the cooling light, i.e., the excitation fraction, to study the cause of its large magnitude.
We found that $\gamma$ depends strongly on the existence of pumping laser beams, particularly at the wavelength of $513\,\mathrm{nm}$ and on the power of the cooling-laser beams, as shown in Fig.~\ref{fig:1-body}.
This could be explained by the assumption of fractional leakage from upper cooling level $\ket{e3}$ to $\ket{m2}$ as one of the main loss mechanism, since the $513$-nm repumper can return the atoms from the metastable state $\ket{m2}$ to the cyclic transition $\ket{m3}\leftrightarrow\ket{e3}$ with an efficiency of $\lesssim 75\,\mathrm{\%}$, where the branching ratio limits the repumping capability.
The finite repumping efficiency causes the residual dependence on the cooling-laser power for the case wherein two repumpers are used.
In addition, the repumping seems to work well even for a small cooling-laser power. This suggests that other decay paths exist from the lower metastable state $\ket{m3}$. Since an even-parity state with $J=11/2$ exists at $1800\,\mathrm{cm^{-1}}$ above $\ket{m3}$ as shown in Fig.~\ref{fig:energylevels}, the excitation from $\ket{m3}$ driven via thermal blackbody radiation and the spontaneous decay to $\ket{m2}$ (and to the other states) may be  participating in the loss mechanism.

In conclusion, we demonstrated a MOT containing up to $1\times10^7$ optically-pumped metastable ${}^{151}\mathrm{Eu}$ atoms.
Measurements of loss from the MOT suggest that, in addition to two-body collisions in the MOT, fractional leakage from the upper cooling level for the MOT limits the number of trapped atoms.

We note that the observed leakage to $\ket{m2}$ will be helpful for pumping the cooled atoms back to the electronic ground state $\ket{g}$ using additional pumping light tuned to the transition $\ket{m2}\leftrightarrow\ket{e1}$ (see Fig.~\ref{fig:energylevels}).
Although the obtained laser-cooled atoms in $\ket{m3,\ F=9}$ have large magnetic moments $\mu=11\,\mathrm{\mu_B}$, a concern in bringing an ensemble of the atoms into quantum degeneracy is that the atoms are in the uppermost energy level in the hyperfine manifold.
Future work will include a pumping the atoms back to $\ket{g}$ and a narrow-line cooling of the atoms using transitions such as $\ket{g} \leftrightarrow 4f^7({}^8\mathrm{S}^{\mathrm{o}})6s6p({}^3\mathrm{P}^\mathrm{o})z\,{}^{10}\mathrm{P}_{9/2}$, where the transition wavelength and the natural linewidth are $687\,\mathrm{nm}$ and $97(2)\,\mathrm{kHz}$ \cite{Fechner1987}, respectively.

We thank S.~Taga and K.~Nishida for experimental assistance.
This study was supported by JSPS KAKENHI (Grants No.~JP16K13856 and No.~JP17J06179); the Murata Science Foundation, and the Research Foundation for Opto-Science and Technology.
One of us (Y.M.) acknowledges partial support from the Japan Society for the Promotion of Science.

\bibliography{mybib}

\begin{thebibliography}{35}%
\makeatletter
\providecommand \@ifxundefined [1]{%
 \@ifx{#1\undefined}
}%
\providecommand \@ifnum [1]{%
 \ifnum #1\expandafter \@firstoftwo
 \else \expandafter \@secondoftwo
 \fi
}%
\providecommand \@ifx [1]{%
 \ifx #1\expandafter \@firstoftwo
 \else \expandafter \@secondoftwo
 \fi
}%
\providecommand \natexlab [1]{#1}%
\providecommand \enquote  [1]{``#1''}%
\providecommand \bibnamefont  [1]{#1}%
\providecommand \bibfnamefont [1]{#1}%
\providecommand \citenamefont [1]{#1}%
\providecommand \href@noop [0]{\@secondoftwo}%
\providecommand \href [0]{\begingroup \@sanitize@url \@href}%
\providecommand \@href[1]{\@@startlink{#1}\@@href}%
\providecommand \@@href[1]{\endgroup#1\@@endlink}%
\providecommand \@sanitize@url [0]{\catcode `\\12\catcode `\$12\catcode
  `\&12\catcode `\#12\catcode `\^12\catcode `\_12\catcode `\%12\relax}%
\providecommand \@@startlink[1]{}%
\providecommand \@@endlink[0]{}%
\providecommand \url  [0]{\begingroup\@sanitize@url \@url }%
\providecommand \@url [1]{\endgroup\@href {#1}{\urlprefix }}%
\providecommand \urlprefix  [0]{URL }%
\providecommand \Eprint [0]{\href }%
\providecommand \doibase [0]{http://dx.doi.org/}%
\providecommand \selectlanguage [0]{\@gobble}%
\providecommand \bibinfo  [0]{\@secondoftwo}%
\providecommand \bibfield  [0]{\@secondoftwo}%
\providecommand \translation [1]{[#1]}%
\providecommand \BibitemOpen [0]{}%
\providecommand \bibitemStop [0]{}%
\providecommand \bibitemNoStop [0]{.\EOS\space}%
\providecommand \EOS [0]{\spacefactor3000\relax}%
\providecommand \BibitemShut  [1]{\csname bibitem#1\endcsname}%
\let\auto@bib@innerbib\@empty
\bibitem [{\citenamefont {Bloch}\ \emph {et~al.}(2008)\citenamefont {Bloch},
  \citenamefont {Dalibard},\ and\ \citenamefont {Zwerger}}]{Bloch2008}%
  \BibitemOpen
  \bibfield  {author} {\bibinfo {author} {\bibfnamefont {I.}~\bibnamefont
  {Bloch}}, \bibinfo {author} {\bibfnamefont {J.}~\bibnamefont {Dalibard}}, \
  and\ \bibinfo {author} {\bibfnamefont {W.}~\bibnamefont {Zwerger}},\ }\href
  {\doibase 10.1103/RevModPhys.80.885} {\bibfield  {journal} {\bibinfo
  {journal} {Rev. Mod. Phys.}\ }\textbf {\bibinfo {volume} {80}},\ \bibinfo
  {pages} {885} (\bibinfo {year} {2008})}\BibitemShut {NoStop}%
\bibitem [{\citenamefont {Lukin}(2003)}]{Lukin2003}%
  \BibitemOpen
  \bibfield  {author} {\bibinfo {author} {\bibfnamefont {M.~D.}\ \bibnamefont
  {Lukin}},\ }\href {\doibase 10.1103/RevModPhys.75.457} {\bibfield  {journal}
  {\bibinfo  {journal} {Rev. Mod. Phys.}\ }\textbf {\bibinfo {volume} {75}},\
  \bibinfo {pages} {457} (\bibinfo {year} {2003})}\BibitemShut {NoStop}%
\bibitem [{\citenamefont {Saffman}\ \emph {et~al.}(2010)\citenamefont
  {Saffman}, \citenamefont {Walker},\ and\ \citenamefont
  {M\o{}lmer}}]{Saffman2010}%
  \BibitemOpen
  \bibfield  {author} {\bibinfo {author} {\bibfnamefont {M.}~\bibnamefont
  {Saffman}}, \bibinfo {author} {\bibfnamefont {T.~G.}\ \bibnamefont {Walker}},
  \ and\ \bibinfo {author} {\bibfnamefont {K.}~\bibnamefont {M\o{}lmer}},\
  }\href {\doibase 10.1103/RevModPhys.82.2313} {\bibfield  {journal} {\bibinfo
  {journal} {Rev. Mod. Phys.}\ }\textbf {\bibinfo {volume} {82}},\ \bibinfo
  {pages} {2313} (\bibinfo {year} {2010})}\BibitemShut {NoStop}%
\bibitem [{\citenamefont {Derevianko}\ and\ \citenamefont
  {Katori}(2011)}]{Derevianko2011}%
  \BibitemOpen
  \bibfield  {author} {\bibinfo {author} {\bibfnamefont {A.}~\bibnamefont
  {Derevianko}}\ and\ \bibinfo {author} {\bibfnamefont {H.}~\bibnamefont
  {Katori}},\ }\href {\doibase 10.1103/RevModPhys.83.331} {\bibfield  {journal}
  {\bibinfo  {journal} {Rev. Mod. Phys.}\ }\textbf {\bibinfo {volume} {83}},\
  \bibinfo {pages} {331} (\bibinfo {year} {2011})}\BibitemShut {NoStop}%
\bibitem [{\citenamefont {McClelland}\ and\ \citenamefont
  {Hanssen}(2006)}]{McClelland2006}%
  \BibitemOpen
  \bibfield  {author} {\bibinfo {author} {\bibfnamefont {J.~J.}\ \bibnamefont
  {McClelland}}\ and\ \bibinfo {author} {\bibfnamefont {J.~L.}\ \bibnamefont
  {Hanssen}},\ }\href {\doibase 10.1103/PhysRevLett.96.143005} {\bibfield
  {journal} {\bibinfo  {journal} {Phys. Rev. Lett.}\ }\textbf {\bibinfo
  {volume} {96}},\ \bibinfo {pages} {143005} (\bibinfo {year}
  {2006})}\BibitemShut {NoStop}%
\bibitem [{\citenamefont {Youn}\ \emph {et~al.}(2010)\citenamefont {Youn},
  \citenamefont {Lu}, \citenamefont {Ray},\ and\ \citenamefont
  {Lev}}]{Youn2010}%
  \BibitemOpen
  \bibfield  {author} {\bibinfo {author} {\bibfnamefont {S.~H.}\ \bibnamefont
  {Youn}}, \bibinfo {author} {\bibfnamefont {M.}~\bibnamefont {Lu}}, \bibinfo
  {author} {\bibfnamefont {U.}~\bibnamefont {Ray}}, \ and\ \bibinfo {author}
  {\bibfnamefont {B.~L.}\ \bibnamefont {Lev}},\ }\href {\doibase
  10.1103/PhysRevA.82.043425} {\bibfield  {journal} {\bibinfo  {journal} {Phys.
  Rev. A}\ }\textbf {\bibinfo {volume} {82}},\ \bibinfo {pages} {043425}
  (\bibinfo {year} {2010})}\BibitemShut {NoStop}%
\bibitem [{\citenamefont {Sukachev}\ \emph {et~al.}(2010)\citenamefont
  {Sukachev}, \citenamefont {Sokolov}, \citenamefont {Chebakov}, \citenamefont
  {Akimov}, \citenamefont {Kanorsky}, \citenamefont {Kolachevsky},\ and\
  \citenamefont {Sorokin}}]{Sukachev2010}%
  \BibitemOpen
  \bibfield  {author} {\bibinfo {author} {\bibfnamefont {D.}~\bibnamefont
  {Sukachev}}, \bibinfo {author} {\bibfnamefont {A.}~\bibnamefont {Sokolov}},
  \bibinfo {author} {\bibfnamefont {K.}~\bibnamefont {Chebakov}}, \bibinfo
  {author} {\bibfnamefont {A.}~\bibnamefont {Akimov}}, \bibinfo {author}
  {\bibfnamefont {S.}~\bibnamefont {Kanorsky}}, \bibinfo {author}
  {\bibfnamefont {N.}~\bibnamefont {Kolachevsky}}, \ and\ \bibinfo {author}
  {\bibfnamefont {V.}~\bibnamefont {Sorokin}},\ }\href {\doibase
  10.1103/PhysRevA.82.011405} {\bibfield  {journal} {\bibinfo  {journal} {Phys.
  Rev. A}\ }\textbf {\bibinfo {volume} {82}},\ \bibinfo {pages} {011405}
  (\bibinfo {year} {2010})}\BibitemShut {NoStop}%
\bibitem [{\citenamefont {Miao}\ \emph {et~al.}(2014)\citenamefont {Miao},
  \citenamefont {Hostetter}, \citenamefont {Stratis},\ and\ \citenamefont
  {Saffman}}]{Miao2014}%
  \BibitemOpen
  \bibfield  {author} {\bibinfo {author} {\bibfnamefont {J.}~\bibnamefont
  {Miao}}, \bibinfo {author} {\bibfnamefont {J.}~\bibnamefont {Hostetter}},
  \bibinfo {author} {\bibfnamefont {G.}~\bibnamefont {Stratis}}, \ and\
  \bibinfo {author} {\bibfnamefont {M.}~\bibnamefont {Saffman}},\ }\href
  {\doibase 10.1103/PhysRevA.89.041401} {\bibfield  {journal} {\bibinfo
  {journal} {Phys. Rev. A}\ }\textbf {\bibinfo {volume} {89}},\ \bibinfo
  {pages} {041401} (\bibinfo {year} {2014})}\BibitemShut {NoStop}%
\bibitem [{\citenamefont {Lahaye}\ \emph {et~al.}(2008)\citenamefont {Lahaye},
  \citenamefont {Metz}, \citenamefont {Fr\"ohlich}, \citenamefont {Koch},
  \citenamefont {Meister}, \citenamefont {Griesmaier}, \citenamefont {Pfau},
  \citenamefont {Saito}, \citenamefont {Kawaguchi},\ and\ \citenamefont
  {Ueda}}]{Lahaye2008}%
  \BibitemOpen
  \bibfield  {author} {\bibinfo {author} {\bibfnamefont {T.}~\bibnamefont
  {Lahaye}}, \bibinfo {author} {\bibfnamefont {J.}~\bibnamefont {Metz}},
  \bibinfo {author} {\bibfnamefont {B.}~\bibnamefont {Fr\"ohlich}}, \bibinfo
  {author} {\bibfnamefont {T.}~\bibnamefont {Koch}}, \bibinfo {author}
  {\bibfnamefont {M.}~\bibnamefont {Meister}}, \bibinfo {author} {\bibfnamefont
  {A.}~\bibnamefont {Griesmaier}}, \bibinfo {author} {\bibfnamefont
  {T.}~\bibnamefont {Pfau}}, \bibinfo {author} {\bibfnamefont {H.}~\bibnamefont
  {Saito}}, \bibinfo {author} {\bibfnamefont {Y.}~\bibnamefont {Kawaguchi}}, \
  and\ \bibinfo {author} {\bibfnamefont {M.}~\bibnamefont {Ueda}},\ }\href
  {\doibase 10.1103/PhysRevLett.101.080401} {\bibfield  {journal} {\bibinfo
  {journal} {Phys. Rev. Lett.}\ }\textbf {\bibinfo {volume} {101}},\ \bibinfo
  {pages} {080401} (\bibinfo {year} {2008})}\BibitemShut {NoStop}%
\bibitem [{\citenamefont {Ferrier-Barbut}\ \emph {et~al.}(2016)\citenamefont
  {Ferrier-Barbut}, \citenamefont {Kadau}, \citenamefont {Schmitt},
  \citenamefont {Wenzel},\ and\ \citenamefont {Pfau}}]{Ferrier-Barbut2016}%
  \BibitemOpen
  \bibfield  {author} {\bibinfo {author} {\bibfnamefont {I.}~\bibnamefont
  {Ferrier-Barbut}}, \bibinfo {author} {\bibfnamefont {H.}~\bibnamefont
  {Kadau}}, \bibinfo {author} {\bibfnamefont {M.}~\bibnamefont {Schmitt}},
  \bibinfo {author} {\bibfnamefont {M.}~\bibnamefont {Wenzel}}, \ and\ \bibinfo
  {author} {\bibfnamefont {T.}~\bibnamefont {Pfau}},\ }\href {\doibase
  10.1103/PhysRevLett.116.215301} {\bibfield  {journal} {\bibinfo  {journal}
  {Phys. Rev. Lett.}\ }\textbf {\bibinfo {volume} {116}},\ \bibinfo {pages}
  {215301} (\bibinfo {year} {2016})}\BibitemShut {NoStop}%
\bibitem [{\citenamefont {Chomaz}\ \emph {et~al.}(2016)\citenamefont {Chomaz},
  \citenamefont {Baier}, \citenamefont {Petter}, \citenamefont {Mark},
  \citenamefont {W\"achtler}, \citenamefont {Santos},\ and\ \citenamefont
  {Ferlaino}}]{Chomaz2016a}%
  \BibitemOpen
  \bibfield  {author} {\bibinfo {author} {\bibfnamefont {L.}~\bibnamefont
  {Chomaz}}, \bibinfo {author} {\bibfnamefont {S.}~\bibnamefont {Baier}},
  \bibinfo {author} {\bibfnamefont {D.}~\bibnamefont {Petter}}, \bibinfo
  {author} {\bibfnamefont {M.~J.}\ \bibnamefont {Mark}}, \bibinfo {author}
  {\bibfnamefont {F.}~\bibnamefont {W\"achtler}}, \bibinfo {author}
  {\bibfnamefont {L.}~\bibnamefont {Santos}}, \ and\ \bibinfo {author}
  {\bibfnamefont {F.}~\bibnamefont {Ferlaino}},\ }\href {\doibase
  10.1103/PhysRevX.6.041039} {\bibfield  {journal} {\bibinfo  {journal} {Phys.
  Rev. X}\ }\textbf {\bibinfo {volume} {6}},\ \bibinfo {pages} {041039}
  (\bibinfo {year} {2016})}\BibitemShut {NoStop}%
\bibitem [{\citenamefont {Kadau}\ \emph {et~al.}(2016)\citenamefont {Kadau},
  \citenamefont {Schmitt}, \citenamefont {Wenzel}, \citenamefont {Wink},
  \citenamefont {Maier}, \citenamefont {Ferrier-Barbut},\ and\ \citenamefont
  {Pfau}}]{Kadau2016}%
  \BibitemOpen
  \bibfield  {author} {\bibinfo {author} {\bibfnamefont {H.}~\bibnamefont
  {Kadau}}, \bibinfo {author} {\bibfnamefont {M.}~\bibnamefont {Schmitt}},
  \bibinfo {author} {\bibfnamefont {M.}~\bibnamefont {Wenzel}}, \bibinfo
  {author} {\bibfnamefont {C.}~\bibnamefont {Wink}}, \bibinfo {author}
  {\bibfnamefont {T.}~\bibnamefont {Maier}}, \bibinfo {author} {\bibfnamefont
  {I.}~\bibnamefont {Ferrier-Barbut}}, \ and\ \bibinfo {author} {\bibfnamefont
  {T.}~\bibnamefont {Pfau}},\ }\href
  {http://www.nature.com/doifinder/10.1038/nature16485} {\bibfield  {journal}
  {\bibinfo  {journal} {Nature}\ }\textbf {\bibinfo {volume} {530}},\ \bibinfo
  {pages} {194} (\bibinfo {year} {2016})}\BibitemShut {NoStop}%
\bibitem [{\citenamefont {Saffman}(2016)}]{Saffman2016}%
  \BibitemOpen
  \bibfield  {author} {\bibinfo {author} {\bibfnamefont {M.}~\bibnamefont
  {Saffman}},\ }\href {http://stacks.iop.org/0953-4075/49/i=20/a=202001}
  {\bibfield  {journal} {\bibinfo  {journal} {Journal of Physics B: Atomic,
  Molecular and Optical Physics}\ }\textbf {\bibinfo {volume} {49}},\ \bibinfo
  {pages} {202001} (\bibinfo {year} {2016})}\BibitemShut {NoStop}%
\bibitem [{\citenamefont {Maier}\ \emph {et~al.}(2015)\citenamefont {Maier},
  \citenamefont {Kadau}, \citenamefont {Schmitt}, \citenamefont {Wenzel},
  \citenamefont {Ferrier-Barbut}, \citenamefont {Pfau}, \citenamefont {Frisch},
  \citenamefont {Baier}, \citenamefont {Aikawa}, \citenamefont {Chomaz},
  \citenamefont {Mark}, \citenamefont {Ferlaino}, \citenamefont {Makrides},
  \citenamefont {Tiesinga}, \citenamefont {Petrov},\ and\ \citenamefont
  {Kotochigova}}]{Maier2015}%
  \BibitemOpen
  \bibfield  {author} {\bibinfo {author} {\bibfnamefont {T.}~\bibnamefont
  {Maier}}, \bibinfo {author} {\bibfnamefont {H.}~\bibnamefont {Kadau}},
  \bibinfo {author} {\bibfnamefont {M.}~\bibnamefont {Schmitt}}, \bibinfo
  {author} {\bibfnamefont {M.}~\bibnamefont {Wenzel}}, \bibinfo {author}
  {\bibfnamefont {I.}~\bibnamefont {Ferrier-Barbut}}, \bibinfo {author}
  {\bibfnamefont {T.}~\bibnamefont {Pfau}}, \bibinfo {author} {\bibfnamefont
  {A.}~\bibnamefont {Frisch}}, \bibinfo {author} {\bibfnamefont
  {S.}~\bibnamefont {Baier}}, \bibinfo {author} {\bibfnamefont
  {K.}~\bibnamefont {Aikawa}}, \bibinfo {author} {\bibfnamefont
  {L.}~\bibnamefont {Chomaz}}, \bibinfo {author} {\bibfnamefont {M.~J.}\
  \bibnamefont {Mark}}, \bibinfo {author} {\bibfnamefont {F.}~\bibnamefont
  {Ferlaino}}, \bibinfo {author} {\bibfnamefont {C.}~\bibnamefont {Makrides}},
  \bibinfo {author} {\bibfnamefont {E.}~\bibnamefont {Tiesinga}}, \bibinfo
  {author} {\bibfnamefont {A.}~\bibnamefont {Petrov}}, \ and\ \bibinfo {author}
  {\bibfnamefont {S.}~\bibnamefont {Kotochigova}},\ }\href {\doibase
  10.1103/PhysRevX.5.041029} {\bibfield  {journal} {\bibinfo  {journal} {Phys.
  Rev. X}\ }\textbf {\bibinfo {volume} {5}},\ \bibinfo {pages} {041029}
  (\bibinfo {year} {2015})}\BibitemShut {NoStop}%
\bibitem [{\citenamefont {Burdick}\ \emph {et~al.}(2016)\citenamefont
  {Burdick}, \citenamefont {Sykes}, \citenamefont {Tang},\ and\ \citenamefont
  {Lev}}]{Nathaniel2016}%
  \BibitemOpen
  \bibfield  {author} {\bibinfo {author} {\bibfnamefont {N.~Q.}\ \bibnamefont
  {Burdick}}, \bibinfo {author} {\bibfnamefont {A.~G.}\ \bibnamefont {Sykes}},
  \bibinfo {author} {\bibfnamefont {Y.}~\bibnamefont {Tang}}, \ and\ \bibinfo
  {author} {\bibfnamefont {B.~L.}\ \bibnamefont {Lev}},\ }\href
  {http://stacks.iop.org/1367-2630/18/i=11/a=113004} {\bibfield  {journal}
  {\bibinfo  {journal} {New Journal of Physics}\ }\textbf {\bibinfo {volume}
  {18}},\ \bibinfo {pages} {113004} (\bibinfo {year} {2016})}\BibitemShut
  {NoStop}%
\bibitem [{\citenamefont {Suleimanov}(2010)}]{Suleimanov2010}%
  \BibitemOpen
  \bibfield  {author} {\bibinfo {author} {\bibfnamefont {Y.~V.}\ \bibnamefont
  {Suleimanov}},\ }\href {\doibase 10.1103/PhysRevA.81.022701} {\bibfield
  {journal} {\bibinfo  {journal} {Phys. Rev. A}\ }\textbf {\bibinfo {volume}
  {81}},\ \bibinfo {pages} {022701} (\bibinfo {year} {2010})}\BibitemShut
  {NoStop}%
\bibitem [{\citenamefont {Buchachenko}\ \emph {et~al.}(2009)\citenamefont
  {Buchachenko}, \citenamefont {Cha{\l}asi{\'n}ski},\ and\ \citenamefont
  {Szcz{\k{e}}{\'s}niak}}]{Alexei2009}%
  \BibitemOpen
  \bibfield  {author} {\bibinfo {author} {\bibfnamefont {A.~A.}\ \bibnamefont
  {Buchachenko}}, \bibinfo {author} {\bibfnamefont {G.}~\bibnamefont
  {Cha{\l}asi{\'n}ski}}, \ and\ \bibinfo {author} {\bibfnamefont {M.~M.}\
  \bibnamefont {Szcz{\k{e}}{\'s}niak}},\ }\href {\doibase 10.1063/1.3282332}
  {\bibfield  {journal} {\bibinfo  {journal} {The Journal of Chemical Physics}\
  }\textbf {\bibinfo {volume} {131}},\ \bibinfo {pages} {241102} (\bibinfo
  {year} {2009})}\BibitemShut {NoStop}%
\bibitem [{\citenamefont {Carr}\ \emph {et~al.}(2009)\citenamefont {Carr},
  \citenamefont {DeMille}, \citenamefont {Krems},\ and\ \citenamefont
  {Ye}}]{carr2009}%
  \BibitemOpen
  \bibfield  {author} {\bibinfo {author} {\bibfnamefont {L.~D.}\ \bibnamefont
  {Carr}}, \bibinfo {author} {\bibfnamefont {D.}~\bibnamefont {DeMille}},
  \bibinfo {author} {\bibfnamefont {R.~V.}\ \bibnamefont {Krems}}, \ and\
  \bibinfo {author} {\bibfnamefont {J.}~\bibnamefont {Ye}},\ }\href
  {https://doi.org/10.1088/1367-2630/11/5/055049} {\bibfield  {journal}
  {\bibinfo  {journal} {New Journal of Physics}\ }\textbf {\bibinfo {volume}
  {11}},\ \bibinfo {pages} {055049} (\bibinfo {year} {2009})}\BibitemShut
  {NoStop}%
\bibitem [{\citenamefont {Tomza}(2014)}]{Tomza2014}%
  \BibitemOpen
  \bibfield  {author} {\bibinfo {author} {\bibfnamefont {M.}~\bibnamefont
  {Tomza}},\ }\href {\doibase 10.1103/PhysRevA.90.022514} {\bibfield  {journal}
  {\bibinfo  {journal} {Phys. Rev. A}\ }\textbf {\bibinfo {volume} {90}},\
  \bibinfo {pages} {022514} (\bibinfo {year} {2014})}\BibitemShut {NoStop}%
\bibitem [{\citenamefont {Papoular}\ \emph {et~al.}(2010)\citenamefont
  {Papoular}, \citenamefont {Shlyapnikov},\ and\ \citenamefont
  {Dalibard}}]{Papoular2010}%
  \BibitemOpen
  \bibfield  {author} {\bibinfo {author} {\bibfnamefont {D.~J.}\ \bibnamefont
  {Papoular}}, \bibinfo {author} {\bibfnamefont {G.~V.}\ \bibnamefont
  {Shlyapnikov}}, \ and\ \bibinfo {author} {\bibfnamefont {J.}~\bibnamefont
  {Dalibard}},\ }\href {\doibase 10.1103/PhysRevA.81.041603} {\bibfield
  {journal} {\bibinfo  {journal} {Phys. Rev. A}\ }\textbf {\bibinfo {volume}
  {81}},\ \bibinfo {pages} {041603} (\bibinfo {year} {2010})}\BibitemShut
  {NoStop}%
\bibitem [{\citenamefont {Hanna}\ \emph {et~al.}(2010)\citenamefont {Hanna},
  \citenamefont {Tiesinga},\ and\ \citenamefont {Julienne}}]{Hanna2010}%
  \BibitemOpen
  \bibfield  {author} {\bibinfo {author} {\bibfnamefont {T.~M.}\ \bibnamefont
  {Hanna}}, \bibinfo {author} {\bibfnamefont {E.}~\bibnamefont {Tiesinga}}, \
  and\ \bibinfo {author} {\bibfnamefont {P.~S.}\ \bibnamefont {Julienne}},\
  }\href {\doibase 10.1088/1367-2630/12/8/083031} {\bibfield  {journal}
  {\bibinfo  {journal} {New Journal of Physics}\ }\textbf {\bibinfo {volume}
  {12}},\ \bibinfo {pages} {083031} (\bibinfo {year} {2010})}\BibitemShut
  {NoStop}%
\bibitem [{\citenamefont {Yi}\ and\ \citenamefont {Pu}(2006)}]{Yi2006}%
  \BibitemOpen
  \bibfield  {author} {\bibinfo {author} {\bibfnamefont {S.}~\bibnamefont
  {Yi}}\ and\ \bibinfo {author} {\bibfnamefont {H.}~\bibnamefont {Pu}},\ }\href
  {\doibase 10.1103/PhysRevLett.97.020401} {\bibfield  {journal} {\bibinfo
  {journal} {Phys. Rev. Lett.}\ }\textbf {\bibinfo {volume} {97}},\ \bibinfo
  {pages} {020401} (\bibinfo {year} {2006})}\BibitemShut {NoStop}%
\bibitem [{\citenamefont {Kawaguchi}\ \emph {et~al.}(2006)\citenamefont
  {Kawaguchi}, \citenamefont {Saito},\ and\ \citenamefont
  {Ueda}}]{Kawaguchi2006}%
  \BibitemOpen
  \bibfield  {author} {\bibinfo {author} {\bibfnamefont {Y.}~\bibnamefont
  {Kawaguchi}}, \bibinfo {author} {\bibfnamefont {H.}~\bibnamefont {Saito}}, \
  and\ \bibinfo {author} {\bibfnamefont {M.}~\bibnamefont {Ueda}},\ }\href
  {\doibase 10.1103/PhysRevLett.97.130404} {\bibfield  {journal} {\bibinfo
  {journal} {Phys. Rev. Lett.}\ }\textbf {\bibinfo {volume} {97}},\ \bibinfo
  {pages} {130404} (\bibinfo {year} {2006})}\BibitemShut {NoStop}%
\bibitem [{\citenamefont {Takahashi}\ \emph {et~al.}(2007)\citenamefont
  {Takahashi}, \citenamefont {Ghosh}, \citenamefont {Mizushima},\ and\
  \citenamefont {Machida}}]{Takahashi2007}%
  \BibitemOpen
  \bibfield  {author} {\bibinfo {author} {\bibfnamefont {M.}~\bibnamefont
  {Takahashi}}, \bibinfo {author} {\bibfnamefont {S.}~\bibnamefont {Ghosh}},
  \bibinfo {author} {\bibfnamefont {T.}~\bibnamefont {Mizushima}}, \ and\
  \bibinfo {author} {\bibfnamefont {K.}~\bibnamefont {Machida}},\ }\href
  {\doibase 10.1103/PhysRevLett.98.260403} {\bibfield  {journal} {\bibinfo
  {journal} {Phys. Rev. Lett.}\ }\textbf {\bibinfo {volume} {98}},\ \bibinfo
  {pages} {260403} (\bibinfo {year} {2007})}\BibitemShut {NoStop}%
\bibitem [{\citenamefont {Pasquiou}\ \emph {et~al.}(2011)\citenamefont
  {Pasquiou}, \citenamefont {Mar\'echal}, \citenamefont {Bismut}, \citenamefont
  {Pedri}, \citenamefont {Vernac}, \citenamefont {Gorceix},\ and\ \citenamefont
  {Laburthe-Tolra}}]{Pasquiou2011}%
  \BibitemOpen
  \bibfield  {author} {\bibinfo {author} {\bibfnamefont {B.}~\bibnamefont
  {Pasquiou}}, \bibinfo {author} {\bibfnamefont {E.}~\bibnamefont
  {Mar\'echal}}, \bibinfo {author} {\bibfnamefont {G.}~\bibnamefont {Bismut}},
  \bibinfo {author} {\bibfnamefont {P.}~\bibnamefont {Pedri}}, \bibinfo
  {author} {\bibfnamefont {L.}~\bibnamefont {Vernac}}, \bibinfo {author}
  {\bibfnamefont {O.}~\bibnamefont {Gorceix}}, \ and\ \bibinfo {author}
  {\bibfnamefont {B.}~\bibnamefont {Laburthe-Tolra}},\ }\href {\doibase
  10.1103/PhysRevLett.106.255303} {\bibfield  {journal} {\bibinfo  {journal}
  {Phys. Rev. Lett.}\ }\textbf {\bibinfo {volume} {106}},\ \bibinfo {pages}
  {255303} (\bibinfo {year} {2011})}\BibitemShut {NoStop}%
\bibitem [{\citenamefont {Martin}\ \emph {et~al.}(1978)\citenamefont {Martin},
  \citenamefont {Zalubas},\ and\ \citenamefont {Hagan}}]{Martin1978}%
  \BibitemOpen
  \bibfield  {author} {\bibinfo {author} {\bibfnamefont {W.~C.}\ \bibnamefont
  {Martin}}, \bibinfo {author} {\bibfnamefont {R.}~\bibnamefont {Zalubas}}, \
  and\ \bibinfo {author} {\bibfnamefont {L.}~\bibnamefont {Hagan}},\
  }\href@noop {} {\bibfield  {journal} {\bibinfo  {journal} {National Standard
  Reference Data Series Vol. 60 (NBS, Washington, D.C., 1978)}\ } (\bibinfo
  {year} {1978})}\BibitemShut {NoStop}%
\bibitem [{\citenamefont {Hartog}\ \emph {et~al.}(2002)\citenamefont {Hartog},
  \citenamefont {Wickliffe},\ and\ \citenamefont {Lawler}}]{Hartog2002}%
  \BibitemOpen
  \bibfield  {author} {\bibinfo {author} {\bibfnamefont {E.~A.~D.}\
  \bibnamefont {Hartog}}, \bibinfo {author} {\bibfnamefont {M.~E.}\
  \bibnamefont {Wickliffe}}, \ and\ \bibinfo {author} {\bibfnamefont {J.~E.}\
  \bibnamefont {Lawler}},\ }\href {\doibase 10.1086/340039} {\bibfield
  {journal} {\bibinfo  {journal} {The Astrophysical Journal Supplement Series}\
  }\textbf {\bibinfo {volume} {141}},\ \bibinfo {pages} {255} (\bibinfo {year}
  {2002})}\BibitemShut {NoStop}%
\bibitem [{\citenamefont {Brand}\ \emph {et~al.}(1981)\citenamefont {Brand},
  \citenamefont {Pfeufer},\ and\ \citenamefont {Steudel}}]{Brand1981}%
  \BibitemOpen
  \bibfield  {author} {\bibinfo {author} {\bibfnamefont {H.}~\bibnamefont
  {Brand}}, \bibinfo {author} {\bibfnamefont {V.}~\bibnamefont {Pfeufer}}, \
  and\ \bibinfo {author} {\bibfnamefont {A.}~\bibnamefont {Steudel}},\ }\href
  {\doibase 10.1007/BF01414259} {\bibfield  {journal} {\bibinfo  {journal}
  {Zeitschrift f{\"u}r Physik A Atoms and Nuclei}\ }\textbf {\bibinfo {volume}
  {302}},\ \bibinfo {pages} {291} (\bibinfo {year} {1981})}\BibitemShut
  {NoStop}%
\bibitem [{\citenamefont {Miyazawa}\ \emph {et~al.}(2017)\citenamefont
  {Miyazawa}, \citenamefont {Inoue}, \citenamefont {Nishida}, \citenamefont
  {Hosoya},\ and\ \citenamefont {Kozuma}}]{Miyazawa2017}%
  \BibitemOpen
  \bibfield  {author} {\bibinfo {author} {\bibfnamefont {Y.}~\bibnamefont
  {Miyazawa}}, \bibinfo {author} {\bibfnamefont {R.}~\bibnamefont {Inoue}},
  \bibinfo {author} {\bibfnamefont {K.}~\bibnamefont {Nishida}}, \bibinfo
  {author} {\bibfnamefont {T.}~\bibnamefont {Hosoya}}, \ and\ \bibinfo {author}
  {\bibfnamefont {M.}~\bibnamefont {Kozuma}},\ }\href {\doibase
  https://doi.org/10.1016/j.optcom.2017.01.059} {\bibfield  {journal} {\bibinfo
   {journal} {Optics Communications}\ }\textbf {\bibinfo {volume} {392}},\
  \bibinfo {pages} {171 } (\bibinfo {year} {2017})}\BibitemShut {NoStop}%
\bibitem [{\citenamefont {Lett}\ \emph {et~al.}(1989)\citenamefont {Lett},
  \citenamefont {Phillips}, \citenamefont {Rolston}, \citenamefont {Tanner},
  \citenamefont {Watts},\ and\ \citenamefont {Westbrook}}]{Lett1989}%
  \BibitemOpen
  \bibfield  {author} {\bibinfo {author} {\bibfnamefont {P.~D.}\ \bibnamefont
  {Lett}}, \bibinfo {author} {\bibfnamefont {W.~D.}\ \bibnamefont {Phillips}},
  \bibinfo {author} {\bibfnamefont {S.~L.}\ \bibnamefont {Rolston}}, \bibinfo
  {author} {\bibfnamefont {C.~E.}\ \bibnamefont {Tanner}}, \bibinfo {author}
  {\bibfnamefont {R.~N.}\ \bibnamefont {Watts}}, \ and\ \bibinfo {author}
  {\bibfnamefont {C.~I.}\ \bibnamefont {Westbrook}},\ }\href {\doibase
  10.1364/JOSAB.6.002084} {\bibfield  {journal} {\bibinfo  {journal} {J. Opt.
  Soc. Am. B}\ }\textbf {\bibinfo {volume} {6}},\ \bibinfo {pages} {2084}
  (\bibinfo {year} {1989})}\BibitemShut {NoStop}%
\bibitem [{\citenamefont {Lu}\ \emph {et~al.}(2010)\citenamefont {Lu},
  \citenamefont {Youn},\ and\ \citenamefont {Lev}}]{Lu2010}%
  \BibitemOpen
  \bibfield  {author} {\bibinfo {author} {\bibfnamefont {M.}~\bibnamefont
  {Lu}}, \bibinfo {author} {\bibfnamefont {S.~H.}\ \bibnamefont {Youn}}, \ and\
  \bibinfo {author} {\bibfnamefont {B.~L.}\ \bibnamefont {Lev}},\ }\href
  {\doibase 10.1103/PhysRevLett.104.063001} {\bibfield  {journal} {\bibinfo
  {journal} {Phys. Rev. Lett.}\ }\textbf {\bibinfo {volume} {104}},\ \bibinfo
  {pages} {063001} (\bibinfo {year} {2010})}\BibitemShut {NoStop}%
\bibitem [{\citenamefont {Vassen}\ \emph {et~al.}(2012)\citenamefont {Vassen},
  \citenamefont {Cohen-Tannoudji}, \citenamefont {Leduc}, \citenamefont
  {Boiron}, \citenamefont {Westbrook}, \citenamefont {Truscott}, \citenamefont
  {Baldwin}, \citenamefont {Birkl}, \citenamefont {Cancio},\ and\ \citenamefont
  {Trippenbach}}]{vassen2012}%
  \BibitemOpen
  \bibfield  {author} {\bibinfo {author} {\bibfnamefont {W.}~\bibnamefont
  {Vassen}}, \bibinfo {author} {\bibfnamefont {C.}~\bibnamefont
  {Cohen-Tannoudji}}, \bibinfo {author} {\bibfnamefont {M.}~\bibnamefont
  {Leduc}}, \bibinfo {author} {\bibfnamefont {D.}~\bibnamefont {Boiron}},
  \bibinfo {author} {\bibfnamefont {C.~I.}\ \bibnamefont {Westbrook}}, \bibinfo
  {author} {\bibfnamefont {A.}~\bibnamefont {Truscott}}, \bibinfo {author}
  {\bibfnamefont {K.}~\bibnamefont {Baldwin}}, \bibinfo {author} {\bibfnamefont
  {G.}~\bibnamefont {Birkl}}, \bibinfo {author} {\bibfnamefont
  {P.}~\bibnamefont {Cancio}}, \ and\ \bibinfo {author} {\bibfnamefont
  {M.}~\bibnamefont {Trippenbach}},\ }\href {\doibase
  10.1103/RevModPhys.84.175} {\bibfield  {journal} {\bibinfo  {journal} {Rev.
  Mod. Phys.}\ }\textbf {\bibinfo {volume} {84}},\ \bibinfo {pages} {175}
  (\bibinfo {year} {2012})}\BibitemShut {NoStop}%
\bibitem [{\citenamefont {Connolly}\ \emph {et~al.}(2010)\citenamefont
  {Connolly}, \citenamefont {Au}, \citenamefont {Doret}, \citenamefont
  {Ketterle},\ and\ \citenamefont {Doyle}}]{Connolly2010}%
  \BibitemOpen
  \bibfield  {author} {\bibinfo {author} {\bibfnamefont {C.~B.}\ \bibnamefont
  {Connolly}}, \bibinfo {author} {\bibfnamefont {Y.~S.}\ \bibnamefont {Au}},
  \bibinfo {author} {\bibfnamefont {S.~C.}\ \bibnamefont {Doret}}, \bibinfo
  {author} {\bibfnamefont {W.}~\bibnamefont {Ketterle}}, \ and\ \bibinfo
  {author} {\bibfnamefont {J.~M.}\ \bibnamefont {Doyle}},\ }\href {\doibase
  10.1103/PhysRevA.81.010702} {\bibfield  {journal} {\bibinfo  {journal} {Phys.
  Rev. A}\ }\textbf {\bibinfo {volume} {81}},\ \bibinfo {pages} {010702}
  (\bibinfo {year} {2010})}\BibitemShut {NoStop}%
\bibitem [{\citenamefont {Kim}\ \emph {et~al.}(1997)\citenamefont {Kim},
  \citenamefont {Friedrich}, \citenamefont {Katz}, \citenamefont {Patterson},
  \citenamefont {Weinstein}, \citenamefont {DeCarvalho},\ and\ \citenamefont
  {Doyle}}]{Kim1997}%
  \BibitemOpen
  \bibfield  {author} {\bibinfo {author} {\bibfnamefont {J.}~\bibnamefont
  {Kim}}, \bibinfo {author} {\bibfnamefont {B.}~\bibnamefont {Friedrich}},
  \bibinfo {author} {\bibfnamefont {D.~P.}\ \bibnamefont {Katz}}, \bibinfo
  {author} {\bibfnamefont {D.}~\bibnamefont {Patterson}}, \bibinfo {author}
  {\bibfnamefont {J.~D.}\ \bibnamefont {Weinstein}}, \bibinfo {author}
  {\bibfnamefont {R.}~\bibnamefont {DeCarvalho}}, \ and\ \bibinfo {author}
  {\bibfnamefont {J.~M.}\ \bibnamefont {Doyle}},\ }\href {\doibase
  10.1103/PhysRevLett.78.3665} {\bibfield  {journal} {\bibinfo  {journal}
  {Phys. Rev. Lett.}\ }\textbf {\bibinfo {volume} {78}},\ \bibinfo {pages}
  {3665} (\bibinfo {year} {1997})}\BibitemShut {NoStop}%
\bibitem [{\citenamefont {Fechner}\ \emph {et~al.}(1987)\citenamefont
  {Fechner}, \citenamefont {Rinkleff},\ and\ \citenamefont
  {Steudel}}]{Fechner1987}%
  \BibitemOpen
  \bibfield  {author} {\bibinfo {author} {\bibfnamefont {B.}~\bibnamefont
  {Fechner}}, \bibinfo {author} {\bibfnamefont {R.-H.}\ \bibnamefont
  {Rinkleff}}, \ and\ \bibinfo {author} {\bibfnamefont {A.}~\bibnamefont
  {Steudel}},\ }\href@noop {} {\bibfield  {journal} {\bibinfo  {journal} {Z.
  Phys. D}\ }\textbf {\bibinfo {volume} {6}},\ \bibinfo {pages} {31} (\bibinfo
  {year} {1987})}\BibitemShut {NoStop}%
\end{thebibliography}%

\end{document}